# A counterexample against the Lesche stability of a generic entropy functional

A. El Kaabouchi[1], Q. A. Wang[1], C.J. Ou[1,2], J.C. Chen[1,3], G. Z. Su[1,3] and A. Le Méhauté[1]

[1]ISMANS, 44, Avenue F.A. Bartholdi, 72000 Le Mans, France
aek@ismans.fr

[2]College of Information Science and Engineering, Huaqiao University,
Quanzhou 362021, People's Republic of China.

[3]Department of Physics and Institute of Theoretical Physics and Astrophysics,
Xiamen University, Xiamen 361005, People's Republic of China

**Abstract:** We provide a counterexample to show that the generic form of entropy $S(p) = \sum_i g(p_i)$ is not always stable against small variation of probability distribution (Lesche stability) even if $g$ is concave function on [0,1] and analytic on ]0,1]. Our conclusion is that the stability of such a generic functional needs more hypotheses on the property of the function $g$, or in other words, the stability of entropy cannot be discussed at this formal stage.

If a physical quantity observable is continuous function of the characteristic variables of motion such as time, configuration, velocity, energy, probability distribution etc., this quantity, and of course its mathematical definition, should undergo smooth variation for the system in smooth motion. Such a condition can be referred to as experimental robustness or observability and can be used to examine the validity of mathematical. An example of such quantity is the entropy which is characteristic of probabilistic uncertainty in stochastic dynamics and considered as continuous function of probability distribution. From this consideration, the mathematical definition of Shannon entropy, Renyi entropy and Tsallis entropy has been reviewed in [1] and [2], in which the robustness was called stability against small perturbation of probability or subsequently Lesche stability after Lesche who initialized the discussion by defining a restrictive uniform continuity criterion [1]. That stability criterion was afterwards used to examine many other quantities including the kappa-entropy [3], the stretched exponential entropy [3], the quantum group entropy [3], the incomplete entropy [4][5] and the escort expectation [6], and has also been extended to a generic forms of entropy $S(p) = \sum_i g(p_i)$ in [7] where the authors advocated for the stability of such a formal definition if $g$ is analytic and concave function of probability distribution $p_i$. This conclusion would be important and very useful if it was well founded. However, we found that a key argument, the calculation from Eq.(2.10) to Eq. (2.15) of [7], would contain some mistakes. As a matter of fact, keeping the Taylor expansion of entropy only to first order in these equations leads to vanishing entropy variation hence invalids the pretended proof of stable $S(p)$. Although this conclusion can be saved by keeping the term of second order in the expansion, the omission of higher order terms cannot be justified. In this letter, we provide a counterexample in order to show that the general conclusion drawn in [7] does not hold. The stability of entropy in that generic form may needs more restrictions on the entropic functional expression.

**Lemma 1.** Let $f$ be a function continuous on an interval $[a,b]$, $g$ a function continuous defined on $f([a,b])$ and $c \in ]a,b[$.

We suppose that:

i) $f$ is of class $C^2$ on $]a,b]$, increasing on $[a,b]$, concave on $[a,c]$ and convex on $[c,b]$;

ii) $g$ is of class $C^2$ on $]f(a), f(b)]$, concave on $[f(a), f(b)]$ increasing on $[f(a), f(c)]$, decreasing on $[f(c), f(b)]$.

Then $g \circ f$ is concave on $[a,b]$.

*Proof*

a) Let $x, y \in [a,c]$ and $\alpha \in [0,1]$. We have

$$f(\alpha x + (1-\alpha)y) \geq \alpha f(x) + (1-\alpha)f(y) \text{ because } f \text{ is concave on } [a,c],$$

and

$$g(f(\alpha x + (1-\alpha)y)) \geq g(\alpha f(x) + (1-\alpha)f(y)) \text{ because } g \text{ is increasing on } [f(a), f(c)].$$

On the other hand,

$$g(\alpha f(x) + (1-\alpha)f(y)) \geq \alpha g(f(x)) + (1-\alpha)g(f(y)) \text{ because } g \text{ is concave on } [f(a), f(b)]$$

It follows that,

$$g(f(\alpha x + (1-\alpha)y)) \geq \alpha g(f(x)) + (1-\alpha)g(f(y)).$$

We deduced thus,

$$(g \circ f)(\alpha x + (1-\alpha)y) \geq \alpha(g \circ f)(x) + (1-\alpha)(g \circ f)(y)$$

Consequently, $g \circ f$ is concave on $[a,c]$ or equivalently $(g \circ f)'' \leq 0$ on $]a,c]$

b)  Let $x, y \in [c,b]$ and $\alpha \in [0,1]$. We have

$$f(\alpha x + (1-\alpha)y) \leq \alpha f(x) + (1-\alpha)f(y) \text{ because } f \text{ is convex on } [c,b],$$

and

$$g(\alpha f(x) + (1-\alpha)f(y)) \leq g(f(\alpha x + (1-\alpha)y)) \text{ because } g \text{ is decreasing on } [f(c), f(b)],$$

On the other hand,

$$g(\alpha f(x) + (1-\alpha)f(y)) \geq \alpha g(f(x)) + (1-\alpha)g(f(y)) \text{ because } g \text{ is concave on } [f(a), f(b)],$$

It follows that,

$$g(f(\alpha x + (1-\alpha)y)) \geq \alpha g(f(x)) + (1-\alpha)g(f(y))$$

We deduce thus

$$(g \circ f)(\alpha x + (1-\alpha)y) \geq \alpha(g \circ f)(x) + (1-\alpha)(g \circ f)(y)$$

Consequently, $g \circ f$ is concave on $[c,b]$ or equivalently $(g \circ f)'' \leq 0$ on $[c,b]$.

c)  By a) $(g \circ f)'' \leq 0$ on $]a,c]$, and b) $(g \circ f)'' \leq 0$ on $[c,b]$, we conclude that $(g \circ f)'' \leq 0$ on $]a,b]$. Thus $g \circ f$ is concave on $[a,b]$.

**Lemma 2.** Let $h$ be the function defined on $[0,1]$ by

$$h(0) = 0 \text{ and } h(x) = \sin\left(\frac{\pi}{1-\ln x}\right) \text{ if } x \neq 0.$$

Then $h$ is concave and verifies $\lim\limits_{x \to +\infty} \dfrac{h\left(\dfrac{1}{x^2}\right)}{h\left(\dfrac{1}{x}\right)} = \dfrac{1}{2}$.

*Proof*

Let $f$ be the function defined on $[0,1]$ by

$$f(0) = 0 \text{ and } f(x) = \frac{\pi}{1-\ln x} \text{ if } x \neq 0,$$

and $g$ be the function defined on $[0, \pi]$ by $g(x) = \sin x$.

We have:

$f$ is continuous on [0,1], $f$ is of class $C^2$ on ]0,1]. For all $x \in ]0,1]$, $f'(x) = \dfrac{\pi}{x(1-\ln x)^2} > 0$, thus $f$ is increasing on [0,1] and we conclude that $f([0,1]) = [f(0), f(1)] = [0, \pi]$.

For all $x \in ]0,1]$, $f''(x) = \dfrac{\pi}{x^2} \dfrac{1+\ln x}{(1-\ln x)^3}$, thus $f''$ is negative on $\left[0, \dfrac{1}{e}\right[$ and $f''$ is positive on $\left[\dfrac{1}{e}, 1\right]$, consequently, $f$ is concave on $\left[0, \dfrac{1}{e}\right]$ and $f$ is convex on $\left[\dfrac{1}{e}, 1\right]$.

On the other hand, $g$ is of class $C^2$ on $[0, \pi]$, $g$ is concave on $[0, \pi]$, increasing on $\left[0, \dfrac{\pi}{2}\right] = \left[f(0), f\left(\dfrac{1}{e}\right)\right]$ and decreasing on $\left[\dfrac{\pi}{2}, \pi\right] = \left[f\left(\dfrac{1}{e}\right), f(1)\right]$. From lemma 1, $h = g \circ f$ is concave on [0,1].

Finally, $\lim\limits_{x \to +\infty} \dfrac{h\left(\dfrac{1}{x^2}\right)}{h\left(\dfrac{1}{x}\right)} = \dfrac{1}{2}$, because we have

$$\dfrac{h\left(\dfrac{1}{x^2}\right)}{h\left(\dfrac{1}{x}\right)} = \dfrac{\sin\left(\dfrac{\pi}{1-\ln(1/x^2)}\right)}{\sin\left(\dfrac{\pi}{1-\ln(1/x)}\right)} = \dfrac{\sin\left(\dfrac{\pi}{1+2\ln x}\right)}{\sin\left(\dfrac{\pi}{1+\ln x}\right)} \underset{\infty}{\approx} \dfrac{\pi}{1+2\ln x} \dfrac{1+\ln x}{\pi}.$$

Thus lemma 2 is proved.

**Lemma 3.** Let $g$ be a function continuous on [0,1] with $g(0) = g(1) = 0$ and $g$ is of class $C^2$ on ]0,1] with $g'' < 0$ on ]0,1].

The entropy associated with $g$ is defined on $E_f = \bigcup\limits_{n \in \mathbf{N}^*} [0,1]^n$ by

$$\forall p = (p_1, \cdots, p_n) \in E_f, \ S(p) = \sum_{i=1}^{n} g(p_i).$$

i) For all $n \in \mathbf{N}^*$, $\sup\left\{S(p), p \in [0,1]^n, \sum\limits_{i=1}^{n} p_i = 1\right\}$ exists and will be denoted by $S_{n,\max}$.

ii) For all $n \in \mathbf{N}^*$, $S_{n,\max} = ng\left(\dfrac{1}{n}\right)$.

*Proof*

i) Let $n \in \mathbf{N}^*$, the function $S_n$ defined on $[0,1]^n$ by : for all $x = (x_1, \cdots, x_n) \in [0,1]^n$,

$S(x) = \sum_{i=1}^{n} g(x_i)$ is continuous on $[0,1]^n$ and $\left\{ x \in [0,1]^n, \sum_{i=1}^{n} x_i = 1 \right\}$ is a compact of $[0,1]^n$

thus $S_{n,\max}$ exists.

ii) Let $n \in \mathbf{N}^*$, we maximize the function $S_n$ with respect $\sum_{i=1}^{n} p_i = 1$.

Using the Lagrange multiplier method, we consider the function $\Phi_n$ defined on $[0,1]^n$ by :

$$\forall p = (p_1, \cdots, p_n) \in [0,1]^n, \ \Phi_n(p) = \sum_{i=1}^{n} g(p_i) - \alpha \left( \sum_{i=1}^{n} p_i - 1 \right).$$

$S_{n,\max} = S_n(p^{(0)})$ where $p^{(0)} = (p_1^{(0)}, p_2^{(0)}, \cdots, p_n^{(0)}) \in [0,1]^n$ and satisfies $\sum_{i=1}^{n} p_i^{(0)} = 1$ for all

$i \in \{1, 2, \cdots, n\}$, $\dfrac{\partial \Phi_n}{\partial x_i}(p^{(0)}) = 0$.

On the other hand, for all $i \in \{1, 2, \cdots, n\}$, $\dfrac{\partial \Phi_n}{\partial x_i}(p) = g'(p_i) - \alpha$.

Because $g'$ is strictly decreasing on $]0,1]$, let $x$ be the unique element of $]0,1]$ such that $g'(x) = \alpha$, we have

$$\forall i \in \{1, 2, \cdots, n\}, \ \dfrac{\partial \Phi_n}{\partial x_i}(p_i^{(0)}) = 0 \Leftrightarrow p_i^{(0)} = x$$

Thus, $\sum_{i=1}^{n} p_i^{(0)} = 1$ implies that $x = \dfrac{1}{n}$.

Consequently, $S_{n,\max} = S_n(p^{(0)}) = \sum_{i=1}^{n} g(p_i^{(0)}) = n g(p_1^{(0)}) = n g\left(\dfrac{1}{n}\right)$.

Thus lemma 3 is proved.

**Definition 4.**

A function $S$ defined on $A = \bigcup_{n \in \mathbf{N}^*} \left\{ p \in [0,1]^n, \sum_{i=1}^{n} p_i = 1 \right\}$ is stable (Lesche stability) if and only if

$$\forall \varepsilon > 0, \exists \delta > 0, \forall N \in \mathbf{N}^*, \forall p, p' \in \left\{ p \in [0,1]^N, \sum_{i=1}^N p_i = 1 \right\},$$

$$\|p - p'\|_1 < \delta \Rightarrow \left| \frac{S(p) - S(p')}{S_{N,\max}} \right| < \varepsilon.$$

**Proposition 5.**

Let $h$ be the function defined on [0,1] by

$$h(0) = 0 \text{ and } h(x) = \sin\left(\frac{\pi}{1 - \ln x}\right) \text{ if } x \neq 0.$$

Then $h$ is concave, $h$ is of class $C^\infty$ on $]0,1]$, continuous on $0$ and $h(0) = h(1) = 0$.

We consider the function $S$ defined on $A = \bigcup_{n \in \mathbf{N}^*} \left\{ p \in [0,1]^n, \sum_{i=1}^n p_i = 1 \right\}$ by

$$\forall n \in \mathbf{N}^*, \forall p = (p_1, \cdots, p_n) \in [0,1]^n \text{ with } \sum_{i=1}^n p_i = 1, \ S(p) = \sum_{i=1}^n g(p_i).$$

Then $S$ is not Lesche stable.

*Proof*

By lemma 2, we can find $N_0 \in \mathbf{N}^*$ such that for all $n \geq N_0$, $\dfrac{n-1}{n} \dfrac{h(1/n^2)}{h(1/n)} > \dfrac{1}{4}$.

For $\varepsilon = \dfrac{1}{8} > 0$, for all $\delta > 0$, there is $N = \max\left( E\left(\dfrac{2}{\delta}\right) + 1, N_0 \right) \in \mathbf{N}^*$, let $p, p'$ defined by

$$p = (p_i)_{i=1,\cdots,N}, \ p_i = \delta_{1i}$$

$$p' = (p'_i)_{i=1,\cdots,N}, \ p'_1 = 1 - \frac{1}{N} + \frac{1}{N^2}, \ p'_i = \frac{1}{N^2} \text{ if } i > 1.$$

We have

$$\sum_{i=1}^N p_i = \sum_{i=1}^W p'_i = 1,$$

$$\|p - p'\|_1 = 1 - \left(1 - \frac{1}{N} + \frac{1}{N^2}\right) + \sum_{i=2}^N \left| 0 - \frac{1}{N^2} \right| = 2\left(\frac{1}{N} - \frac{1}{N^2}\right) \leq \frac{2}{N} < \delta,$$

and

$$\left|\frac{S(p)-S(p')}{S_{N,\max}}\right| = \frac{\left|h(1)-h\left(1-\frac{1}{N}+\frac{1}{N^2}\right)+\sum_{i=2}^{N}\left(h(0)-h\left(\frac{1}{N^2}\right)\right)\right|}{Nh\left(\frac{1}{N}\right)}$$

$$= \frac{\left|-h\left(1-\frac{1}{N}+\frac{1}{N^2}\right)-(N-1)h\left(\frac{1}{N^2}\right)\right|}{Nh\left(\frac{1}{N}\right)}$$

$$= \frac{h\left(1-\frac{1}{N}+\frac{1}{N^2}\right)+(N-1)h\left(\frac{1}{N^2}\right)}{Nh\left(\frac{1}{N}\right)}$$

$$\geq \frac{(N-1)h\left(\frac{1}{N^2}\right)}{Nh\left(\frac{1}{N}\right)} \geq \frac{1}{4} > \varepsilon$$

Consequently, $S$ is not Lesche stable.

In summary, a counterexample is given to show that the entropy $S(p)=\sum_i g(p_i)$ is not always stable or robust even if $g$ is concave function on [0,1] and analytic on ]0,1]. To our opinion, the stability of such a generic form of entropy needs more hypotheses on the property of the function $g$, or the stability of entropy cannot be discussed at this formal stage.

We would like to indicate in passing that the present counterexample can also be used to review the conclusion of [3] about the stability of a generic definition of entropy, given by Eq.(6) of [3] based on maximum entropy principle (maxent). In fact, if one replaces the function $\int_0^{p_i} dt\, f^{-1}(t)$ in that equation by $g(p_i)$ or $f(t)$ by $g^{-1}(t)$ which satisfies the properties of $f(t)$, then the function $B(\bullet)$ defined in Eq.(18) of [3] has a limit value of 1 instead of zero as obtained in [3]. Note however that the eventual failure of establishing stability for this formal entropy has nothing to do with the validity of that definition which we esteem on the contrary a logical and useful proposal based on maxent. The stability problem can be addressed after an entropy is derived from this definition, as has been done in [3] for several functional. This same statement also applies to the definition, proposed by mimicking the second law entropy for reversible process, of a variational entropy (called varentropy) $dS = \overline{dx} - \overline{dx}$ for any single random variable *x*[8]. Varentropy has been shown to be maximized by the distributions used to derive it's functional [8]. For this purpose, it coincides with the entropy of Eq.(6) in [3].

**Acknowledgements**

We thank Professor Abe for valuable discussions. Two of the authors (Chen and Ou) thank the Région des Pays de la Loire of France for the grants N° 2007-6088 and N°2007-11953, Ou also thanks Huaqiao University, People's Republic of China for the Science Research Fund (No. 07BS105). Su thanks ISMANS for the support of his participation in the research works around this subject.